\documentclass[letterpaper]{article}
\usepackage{aaai2027}
\usepackage[hyphens]{url}
\usepackage{graphicx}
\usepackage{natbib}
\usepackage{caption}
\usepackage{booktabs}
\usepackage{amsmath}
\usepackage{amssymb}

\newif\ifarxiv
\arxivfalse
\IfFileExists{arxivmode.tex}{\arxivtrue}{}

\ifarxiv
\fi

\title{Spend Classification Without Leakage: An Evaluation Harness\\
and What It Changed in a Deployed System}

\author{
    Harshit Gupta
}
\affiliations{
    SpendSide\\
    San Francisco, United States\\
    harshitg93@gmail.com
}

\begin{document}
\maketitle

\begin{abstract}
Assigning a standard commodity code to a free-text purchase line is the foundation of
enterprise spend analytics, and the accuracy claimed for it cannot be checked. Published
results are measured on proprietary data, or on private samples of public data under
undocumented protocols, so no two are comparable and the task has no public
benchmark. We build a harness for it: four evaluation protocols over 1.26 million labelled
purchase lines released by two US state governments. We use it
to show that the standard protocol overstates what these systems can do. Enterprises rebuy
the same items continuously, so a random train/test split places matching item
text on both sides of the partition: 60.7\% and 61.2\% of test rows in our two corpora, and
54.7\% and 60.6\% when the match must be byte for byte.
Separating accuracy by whether the test item was already seen shows what that buys. On
California purchase orders the strongest classical baseline scores 56.0\% on repeated text
but \textbf{31.9\% on genuinely novel text}. That 24-point gap widens monotonically with
taxonomy depth. For nearest-neighbour retrieval, sentence embeddings shrink the equivalent
gap only from 23.8 to 19.1 points, and a fine-tuned transformer does not escape it either.
On one corpus the leaky protocol cannot separate that retrieval baseline from the
transformer, while all three leak-free protocols put the transformer 2.3 to 3.6 points
ahead, so the standard split can conceal a real difference between two models rather than
merely flatter both. We then derive an exact upper bound on any classifier
that sees only item text, from cases where identical text carries conflicting codes. It is low enough to reframe what the
field reports, and it is not a single number: 79.4\% at commodity level on one corpus
against 98.4\% on the other, so accuracy figures are not comparable across datasets even in
principle.
The same bound weighted by spend says more about the amount field than about the
labels, which is a check worth running before weighting accuracy by any
spend column. Finally, in a commercial deployment covering 920,927 purchase lines, human
reviewers accepted \textbf{32.4\%} of the system's suggestions on the population our
novel-text estimate describes, pooled across three organisations, and the 95\% interval
29.7--35.2 brackets the benchmark's 31.9\% and 34.8\%. To our knowledge that is the first
external check of a public procurement benchmark against live human judgement. We release the harness, which is in
production use: it moved that deployment off a text-only design and set the abstention
policy it now ships.
\end{abstract}

\section{Introduction}

A spend cube underlies every category strategy, supplier consolidation exercise and
savings programme. It is built by assigning each purchase line a code from a
standard commodity taxonomy. The input is short, abbreviation-dense, human-typed text
(\texttt{DURACELL 2032 LITHIUM BATT 9CT}; \texttt{MISC PLUMBING SUPPLIES BLDG 4}). The
output is a node in a hierarchy such as UNSPSC, which is organised in four levels of two
digits each: Segment, Family, Class, Commodity. Enterprises process $10^6$ to $10^8$ such
lines, so the task is automated or it is not done.

The task was named and framed in 2008 \citep{mukherjee2008}, and procurement scholarship has
since converged on AI as central to the function \citep{guida2023}. Eighteen years later
there is still no public benchmark. Nearly every result we located is on data nobody else
can obtain: Siemens internal spend \citep{mukherjee2008}, proprietary Italian tender data
\citep{moiraghi2024}, a single UK food manufacturer \citep{li2025}, Korea Customs Service
records \citep{lee2024}. A recent qualitative study of vendor offerings \citep{guida2025}
contains no dataset, no baseline and no accuracy figure, and describes the area as largely
unexplored. No two of these results are comparable: each is measured on different data
under an undocumented protocol.

That vacuum has been filled by results that cannot be reconstructed. \citet{singh2024}
evaluates GPT-4 on a sample of the same California purchase orders we use, and the 90\%
headline belongs to the follow-up \citep{singh2025}, at Segment level on a restricted
subset. The problem is not that such numbers are wrong, but that no published result in
this area states its protocol in enough detail to rebuild, so none can be set against
another, or against ours.

This paper is an attempt to put a floor under the question. Our contributions:

\begin{itemize}
\item \textbf{A public evaluation harness.} Four protocols over 1,258,167 usable labelled
purchase lines from two independent US state government sources, two eras and two
procurement channels, which a practitioner runs against their own
corpus (Section~\ref{sec:benchmark}).

\item \textbf{The size of a known leak, in a domain that could not measure it.} Duplicate
text across a split is a studied failure \citep{elangovan2021,kapoor2023} and grouping
splits against it is established practice \citep{peeters2024}; procurement had no
instrument to measure it. A random split puts matching text on both sides for three test
rows in five, and we decompose every baseline's accuracy into already-seen and novel text
(Sections~\ref{sec:data} and~\ref{sec:results}).

\item \textbf{A limit on attainable accuracy that no procurement study reports:} an
annotation-conflict ceiling affecting 47.6\% of rows in one corpus, and the finding that the
amount field one of these corpora publishes will not carry a weighted version of it
(Section~\ref{sec:data}).

\item \textbf{The harness in production use, and an external check on it.} Inside
SpendSide, a commercial spend analytics product in pilot use with three organisations, it
replaced a text-only classifier with a deterministic cascade and set an abstention policy.
What reviewers then accepted matched the leak-free estimate rather than the conventional
one (Sections~\ref{sec:application} and~\ref{sec:deployment}).
\end{itemize}

\section{The Application and Its Path to Deployment}
\label{sec:application}

SpendSide is a commercial spend analytics product, in pilot use with three organisations
that have loaded live purchase data: 920,927 lines in total, spanning five fiscal years.
Classification is not a feature of such a product but its precondition. Category rollups,
supplier analytics and every saving opportunity are computed over assigned codes, so a
classification error does not stay local. It propagates into every number the user sees.

Two properties of the labels differ from the benchmark and from most published work. First,
each organisation is classified into \emph{its own} three-level taxonomy of between 17 and
40 top-level categories, derived from the lines it had already coded and from its parts
master, with an 8-digit UNSPSC code assigned alongside. A deployment inherits whatever
taxonomy its customer already uses; it does not get to pick a clean one, and the taxonomies
are not mutually comparable.

Second, coverage as received was roughly half of all lines. Where it stands now, and how it
got there, is the subject of Section~\ref{sec:deployment}, and it is not what the benchmark
would lead one to expect.

\paragraph{Path to deployment.} The blockers that gate a pilot becoming a paying
deployment are closed: loaded data persists, savings are derived from that data rather
than back-solved from a target, and accept and reject decisions are durable. What remains
is specific. Seven opportunity levers are notional, and stay switched off until data makes
them real. One organisation's performance-driver column is unusable, which leaves only 53
of 12,266 candidate items eligible for like-for-like price modelling. That is a
data-quality blocker, not a modelling one. And all three
tenants were loaded against a January fiscal-year start because that question was never
put to them, which has to be corrected per tenant before their reporting periods mean
anything.

\section{Related Work}

\paragraph{Spend and commodity classification.}
\citet{mukherjee2008} framed the task with off-the-shelf SVM and logistic regression on
Siemens spend data, and their statement of the difficulty (sparse abbreviation-heavy text,
extreme class imbalance, high volume) remains accurate. \citet{moiraghi2024} is the
closest methodological prior work: zero-shot hierarchical classification into the EU Common
Procurement Vocabulary using label descriptions, but trained and evaluated on proprietary
data. \citet{lee2024} assign Harmonized
System codes with expert-validated explanations, \citet{li2025} replicate expert supplier
categorisation at a single firm, and \citet{abdullahi2024} is one of very few works to
classify actual government spend to UNSPSC. None of it is reproducible by a third party.

\paragraph{Hierarchical and extreme classification.}
Techniques for large, structured label spaces are well developed outside procurement.
\citet{zhang2025teleclass} enrich a taxonomy with mined class-indicative features and note
that naive zero-shot prompting degrades in hierarchical settings; \citet{guo2025green}
decompose a 1,114-way NAICS decision along the ontology; \citet{nath2025} generate
hierarchical code sequences; \citet{molfetta2025} target rare classes in a regulatory food
taxonomy; \citet{ortego2026} establish practice on extreme multi-label benchmarks.
Product-taxonomy shared tasks \citep{mwpd2020,rakuten2018} are the closest evaluation
designs to what procurement lacks, and both were last run in 2020.

\paragraph{Evaluation practice, and prior art on leakage.}
Our framing follows \citet{kahler2026}, who compare supervised extreme classification
against lexical and generative baselines on one corpus with multiple metrics and report a
split verdict; that design has not reached a procurement taxonomy. \citet{kapoor2024} argue
that a narrow focus on accuracy without cost has led the community to mistaken conclusions,
and \citet{zhu2025} show that benchmark task-setup flaws can misstate performance by up to
100\% in relative terms and propose a checklist we follow. The leakage mechanism is not
ours either: \citet{elangovan2021} quantify train/test text overlap on public NLP corpora
and show that scores inflate with it, \citet{kapoor2023} give duplicates in datasets a type
of their own in a leakage taxonomy for ML-based science, and \citet{peeters2024} already
make unseen-entity generalisation a benchmark dimension in product matching. What none
supplies is the magnitude in procurement, or a bound derived from the labels themselves.

\section{Harness Construction}
\label{sec:benchmark}

\subsection{Corpora}

We use two independent government sources, chosen because they are genuinely public,
carry buyer-assigned UNSPSC codes, and differ from each other in ways that matter.

\textbf{California DGS purchase orders (\textsc{ca-dgs}).} State purchase orders for
fiscal years 2012--13 through 2014--15, released with no restrictions on public use.
344,504 rows, of which 343,484 are usable. The text field is an ERP item name with a median
length of 19 characters, against a broad label space.

\textbf{Washington State agency Amazon marketplace spend (\textsc{wa-amazon}).} Agency
purchases through a marketplace contract, fiscal years 2018--2026, published per year on
the state open-data portal. 944,699 rows, of which 914,683 are usable. The text field is a
full marketplace product title, typically an order of magnitude longer, against a label
space roughly five times narrower.

The pair is not two samples of one distribution. One is terse ERP free-text over 13,399
commodity codes; the other is verbose merchandising copy over 2,629, which makes
cross-corpus transfer a meaningful question rather than a formality.

\subsection{Label Normalisation and Two Source Defects}

UNSPSC codes are a positional hierarchy, so a short code is a truncated prefix and its
missing depth is trailing zeros. No valid code begins
with a zero, since segments start at 10, and none of the 343,469 and 914,687 source values
that arrive already eight digits long does. Both counts are of source values rather than
corpus rows. A six-digit code such as
\texttt{401728} therefore denotes Class 4017.28 and must be right-padded to
\texttt{40172800}. Left-padding it to \texttt{00401728}, which is what a naive
zero-fill produces, invents a segment that does not exist: few rows in either corpus, but
silently rather than visibly. We reject
codes with fewer than two digits or a segment below 10.

The second defect is temporal. California's \texttt{Purchase Date} cannot order time, so
the temporal protocol keys on its clean \texttt{Fiscal Year} field, and on calendar
quarter of \texttt{Order Date} for \textsc{wa-amazon}.

We derive the four levels by prefix slicing. \textsc{ca-dgs} also ships pre-joined
Class, Family and Segment columns, which lets us check the derivation against the
publisher's own: after the padding correction the two agree on 341,211 of 341,211
comparable rows at all three levels, exactly.

\subsection{Heterogeneous Label Depth}

\ifarxiv
\begin{table}[t]
\centering
\small
\begin{tabular}{llrr}
\toprule
Corpus & Level & Evaluable rows & Labels \\
\midrule
\textsc{ca-dgs} & Segment   & 343,484 (100.0\%) & 56 \\
                & Family    & 343,484 (100.0\%) & 411 \\
                & Class     & 343,484 (100.0\%) & 2,377 \\
                & Commodity & 343,467 (99.995\%) & 13,399 \\
\midrule
\textsc{wa-amazon} & Segment   & 914,683 (100.0\%) & 51 \\
                   & Family    & 870,254 (95.1\%)  & 273 \\
                   & Class     & 757,434 (82.8\%)  & 996 \\
                   & Commodity & 365,085 (39.9\%)  & 2,629 \\
\bottomrule
\end{tabular}
\caption{Rows evaluable at each hierarchy level. Depth is heterogeneous, so the
denominator changes with level and must be stated. The \textsc{ca-dgs} Commodity share is
99.995\% rather than 100\%: 17 rows carry a six-digit code and are evaluable no deeper
than Class. Rounding it to 100.0\% would have retired those rows by presentation, which is
the failure this column exists to prevent.}
\label{tab:evaluable}
\end{table}
\fi

Labels are not all annotated to the same depth, and treating them as though they were
inflates or deflates results depending on the level examined. A label truncated at Class
carries no Commodity information, so it is evaluable at Class and above and not below.
Table~\ref{tab:evaluable} gives the resulting denominators. In
\textsc{wa-amazon} only 39.9\% of rows support Commodity-level evaluation, against 100\% at
Segment. Any single headline accuracy figure for this corpus is therefore measured over an
unstated denominator; we report the denominator at every level.

\subsection{Four Evaluation Protocols}

We define four partitions, each targeting 20\% test, and we retain the flawed one on
purpose. \textsc{temporal} on \textsc{ca-dgs} is the exception: only three fiscal years
exist, so holding out the last gives 33.7\%.

\textbf{\textsc{iid-naive}} is a row-level random shuffle. It leaks repeated items across
the partition by construction. We include it because it approximates what the literature
reports, and the gap between it and \textsc{iid-grouped} is the quantity this harness
exists to measure.

\textbf{\textsc{iid-grouped}} randomises over normalised-text groups, so every row sharing
an item description falls on the same side. Leak-free by construction, and verified so.

\textbf{\textsc{temporal}} trains on earlier periods and tests on later ones, whole periods
at a time. This is what a deployed classifier actually faces.

\textbf{\textsc{supplier}} holds out entire suppliers, which is what onboarding a customer
whose vendors the model has never seen actually faces.

For \textsc{temporal} and \textsc{supplier} we measure residual text overlap but do not
remove it, because items genuinely are rebought and two sellers genuinely do list the same
product. Only \textsc{iid-grouped} claims leak-freedom, and the build asserts it.

\section{What the Data Says Before Any Model}
\label{sec:data}

\subsection{A Random Split Is Mostly a Memory Test}

\ifarxiv
\begin{table}[t]
\centering
\small
\begin{tabular}{lrr}
\toprule
Protocol & \textsc{ca-dgs} & \textsc{wa-amazon} \\
\midrule
\textsc{iid-naive}   & 60.7\% & 61.2\% \\
\textsc{iid-grouped} & 0.0\%  & 0.0\%  \\
\textsc{temporal}    & 51.9\% & 27.2\% \\
\textsc{supplier}    & 45.6\% & 41.9\% \\
\bottomrule
\end{tabular}
\caption{Share of test rows whose normalised item text already appears in training. Under a
row-level shuffle roughly three in five test items have been seen already.}
\label{tab:leakage}
\end{table}
\fi

For each protocol, Table~\ref{tab:leakage} reports the share of test rows
whose item text already appears in training, matched after case-folding, replacing
punctuation with spaces and collapsing whitespace. Under the row-level shuffle it is 60.7\%
and 61.2\%; under the grouped split, nil. That normalisation is not what produces the
effect. Across ten equivalence relations, from byte-identical to an order-insensitive token
set, the row-level share stays within 54.7--61.9\% on \textsc{ca-dgs} and 60.6--61.6\% on
\textsc{wa-amazon}, and no normalisation at all gives the low end of each.

The cause is mundane and unavoidable: organisations rebuy. 54.1\% of \textsc{ca-dgs} rows
and 50.6\% of \textsc{wa-amazon} rows are repeated text, and in \textsc{wa-amazon} 57.6\% of
rows carrying a marketplace product identifier are repeat purchases of the identical
product. A shuffle cannot help but split those repeats.

\subsection{An Annotation Ceiling, and What the Spend Field Does to It}

\ifarxiv
\begin{table}[t]
\centering
\small
\setlength{\tabcolsep}{4pt}
\begin{tabular}{llrrr}
\toprule
Corpus & Level & \% rows in & \multicolumn{2}{c}{Ceiling (\%)} \\
\cmidrule(l){4-5}
       &       & conflict   & count-wtd & spend-wtd \\
\midrule
\textsc{ca-dgs} & Segment   & 40.0 & 89.2 & 98.5 \\
                & Family    & 42.6 & 86.0 & 98.0 \\
                & Class     & 44.6 & 83.5 & 97.8 \\
                & Commodity & 47.6 & \textbf{79.4} & \textbf{97.4} \\
\midrule
\textsc{wa-amazon} & Segment   & 3.5 & 99.2 & 99.2 \\
                   & Family    & 4.2 & 99.1 & 99.0 \\
                   & Class     & 4.7 & 99.0 & 98.8 \\
                   & Commodity & 6.9 & \textbf{98.4} & \textbf{98.4} \\
\bottomrule
\end{tabular}
\caption{Highest accuracy achievable by any classifier that sees only item text. Computed on rows annotated to full commodity depth so that the denominator is constant across levels (343,467 rows for \textsc{ca-dgs}, 365,085 for \textsc{wa-amazon}); a coarse label is a prefix of the same row's fine label, so conflict rises and the ceiling falls with depth, as observed.}
\label{tab:ceiling}
\end{table}
\fi

Identical item text sometimes carries different codes within the same corpus. In
\textsc{ca-dgs}, 15,383 distinct texts (9.8\% of distinct texts) appear under more than
one commodity code, and those texts account for \textbf{47.6\% of all rows}.

This admits an exact bound. Every model we evaluate is a function of item text, so if one
text appears under two codes no such function can be right
on both. The best any text-only classifier can achieve is to predict, for each distinct
text, that text's most frequent label:
\[
\text{ceiling} \;=\; \frac{1}{N}\sum_{t \in \mathcal{T}} \max_{c} \; n(t, c),
\]
for texts $\mathcal{T}$ and label counts $n(t,c)$. It requires no model and no training,
and it holds for every result in Section~\ref{sec:results}.

Table~\ref{tab:ceiling} evaluates it. Two things follow, and the second
one surprised us.

First, the ceiling is severe and it is corpus-specific. On \textsc{ca-dgs} no text-only
classifier can exceed \textbf{79.4\%} at Commodity level. A paper reporting 85\% on terse
ERP text of this kind is not reporting a good model; it is reporting a leak, a different
denominator, or a different corpus. On \textsc{wa-amazon} the ceiling is 98.4\%, because a
full marketplace product title nearly determines its category. The two corpora differ by
19 points in what is even attainable, which is a direct argument against comparing accuracy
figures across datasets, the practice the entire field currently relies on.

Second, weighting the same bound by spend rather than by row gives 97.4\% on
\textsc{ca-dgs} against that 79.4\%, and the 18 points between them are a property of the
amount field rather than of the labels. California's field mixes contract awards with
ordinary line totals, and its median line of \$3,565 sits against a mean of \$438,672.
One commodity code carries 70.4\% of \$150.7 billion over 4,204 rows,
25 lines above \$1 billion carry 32.7\% of it, and the largest single line is \$7.3 billion
of \texttt{Personal Service}. Under spend weights the test set has a Kish effective sample
size of 19.9, which is why no spend-weighted figure in this paper carries an interval.

Those weights land where conflict is impossible. 63.9\% of the money sits on item text
appearing exactly once, where the modal code is the only code, and reweighting the
count-weighted shares by spend while leaving every conflict untouched already reaches
96.0\%. Among text that can conflict, agreement falls as line value rises rather than
improving, so the divergence is not careful coding of expensive lines. The bound drops to
88.2\% winsorised at the 99th percentile and to 79.5\% under log spend, back onto the
count-weighted figure. \textsc{wa-amazon}, whose lines average \$96 and whose effective
sample is 11,488 rather than 19.9, gives 98.4\% both ways. We report the count-weighted
ceiling and read the spend-weighted one as a statement about the field. Checking what an
amount column measures before weighting accuracy by it is the part that transfers.

Some of what any paper reports as model error is disagreement among the human
buyers who assigned the labels \citep{northcutt2021,plank2022,pavlick2019}, though we have
not found it turned into a bound for a procurement corpus. Not all of it is: California's
publisher retained only the first code where a purchase order carried several, so part of
the conflict is a second valid code dropped at source. Read the ceiling as a bound on
single-label accuracy over this data, not a limit on the task.

\section{Baselines and Results}
\label{sec:results}

\subsection{Setup}

We evaluate a most-frequent-label floor, Complement naive Bayes, a nearest class centroid
under cosine similarity, and 1-nearest-neighbour cosine retrieval, all over word-level
TF-IDF ($1$--$2$-grams, sublinear term frequency, $\min$ document frequency 2).

We additionally run 1-NN retrieval and a nearest centroid over
\texttt{all-MiniLM-L6-v2} sentence embeddings, normalised so that cosine is a dot product.
The highest-capacity baseline is a fine-tuned \texttt{distilbert-base-uncased}, 67.0 million
parameters, three epochs per cell with a fresh head over that cell's label space, on the
same splits and the same training cap as every other model.

The main grid caps the vocabulary at 30,000 features at every level, a memory constraint
rather than a tuned choice; Section~\ref{sec:lexical} reports what lifting it costs.

Every configuration is scored for accuracy, macro-$F_1$ and spend-weighted accuracy, and
separately on test rows whose text was already seen in training and on those whose text is
novel. Accuracy answers the operational question, macro-$F_1$ keeps the head of the
distribution from hiding tail collapse, and spend-weighted accuracy is recorded but not
argued from, for the reason Section~\ref{sec:data} gives.

Every setting above is fixed a priori and applied identically in every cell, so nothing is
selected on test data. Training is capped at 200,000 rows per cell, and the decomposition
below is computed against that capped training set rather than the whole partition
Table~\ref{tab:leakage} reports: under \textsc{iid-naive} the seen-text share the models met
is 58.2\% on \textsc{ca-dgs} and 42.3\% to 61.5\% across levels on \textsc{wa-amazon},
against the 60.7\% and 61.2\% of the table. Capping lowers it in every cell, so each model
saw less repetition than its protocol allows and Table~\ref{tab:gap} understates if
anything. The table's two figures cover all test rows rather than one level, which is why
the \textsc{wa-amazon} range reaches above them.
Experiments ran on an Apple M2 with 24\,GB of unified memory, and the appendix records the
software environment.

Each cell is run once. The classical and embedding estimators are deterministic given a
fixed split, and the splits are fixed by seed 20260824, so repeating them would reproduce
each number rather than estimate a variance. Fine-tuning is seeded per cell but its training
kernels are not bit-reproducible, and one run carries no seed-to-seed variance, which the
intervals below therefore omit. Uncertainty is
reported as the Wilson score interval on the test proportion; the largest
half-width in the complete grid is 0.37 points, and 0.60 on the seen and novel subgroups,
so the seen-against-novel gaps below are far larger than sampling error. Because item text repeats, rows are not fully independent on
the leaking protocols, which makes that interval a lower bound on true uncertainty rather
than an exact one.

\subsection{What a Headline Number Is Made Of}

\ifarxiv
\begin{table}[t]
\centering\small
\begin{tabular}{llrrr}
\toprule
Corpus & Model & Seen & Novel & Gap \\
\midrule
\textsc{ca-dgs} & Naive Bayes & 56.0 & 31.9 & \textbf{24.1} \\
 & Centroid & 28.7 & 15.7 & \textbf{13.0} \\
 & 1-NN & 53.0 & 29.2 & \textbf{23.8} \\
\midrule
\textsc{wa-amazon} & Naive Bayes & 84.9 & 69.4 & \textbf{15.5} \\
 & Centroid & 78.6 & 63.9 & \textbf{14.7} \\
 & 1-NN & 95.9 & 73.9 & \textbf{22.0} \\
\bottomrule
\end{tabular}
\caption{Commodity-level accuracy (\%) under the leaky protocol, separated by whether the test row's item text appeared in training. A single headline number averages these two populations and reports neither.}
\label{tab:seenunseen}
\end{table}
\fi

Table~\ref{tab:seenunseen} splits accuracy under the leaking protocol by whether the test
row's item text had been seen in training. The two populations behave nothing alike. On
\textsc{ca-dgs} at Commodity level, Complement naive Bayes scores 56.0\% on repeated text
and \textbf{31.9\% on novel text}; 1-NN scores 53.0\% against 29.2\%. The reported
aggregate, 45.9\% and 43.0\% respectively, is an average over two populations and
describes neither.

The gap widens with taxonomy depth: for naive Bayes it runs 20.6 points at Segment, 21.1 at
Family, 22.5 at Class and 24.1 at Commodity. This is the direction that matters
operationally, because Segment is too coarse to source against and Commodity is the level a
category manager actually needs.

A buyer onboarding a new dataset should expect roughly the novel-text figure, not the
headline. At Commodity level on terse ERP text that is about 32\%.

\subsection{Is This an Artefact of Weak Baselines?}
\label{sec:lexical}

\ifarxiv
\begin{table}[t]
\centering\small
\begin{tabular}{llrrr}
\toprule
Corpus & Representation & Seen & Novel & Gap \\
\midrule
\textsc{ca-dgs} & 1-NN & 53.0 & 29.2 & \textbf{23.8} \\
 & 1-NN (embed.) & 53.9 & 34.8 & \textbf{19.1} \\
\addlinespace
 & Centroid & 28.7 & 15.7 & \textbf{13.0} \\
 & Centroid (embed.) & 32.2 & 22.0 & \textbf{10.2} \\
\addlinespace
\midrule
\textsc{wa-amazon} & 1-NN & 95.9 & 73.9 & \textbf{22.0} \\
 & 1-NN (embed.) & 96.3 & 76.6 & \textbf{19.6} \\
\addlinespace
 & Centroid & 78.6 & 63.9 & \textbf{14.7} \\
 & Centroid (embed.) & 77.3 & 66.7 & \textbf{10.6} \\
\addlinespace
\bottomrule
\end{tabular}
\caption{Commodity-level accuracy (\%) under the leaky protocol, lexical TF-IDF against \texttt{all-MiniLM-L6-v2} sentence embeddings, on identical splits with identical scoring. The seen-against-novel gap does not close under a semantic representation, so it is a property of the task rather than of lexical matching.}
\label{tab:neural}
\end{table}
\fi

There is an obvious objection here, and it is a good one. TF-IDF is a bag of words, so it
matches a repeated item verbatim and has little to fall back on when the words change.
Perhaps the gap measures our choice of baseline rather than the task, and a better
representation or a larger model would generalise across paraphrase and close it.

Neither does. Table~\ref{tab:neural} repeats the decomposition with
\texttt{all-MiniLM-L6-v2} sentence embeddings substituted for TF-IDF, on identical splits,
denominators and scoring code, so the representation is the only thing that
differs. Embeddings do help on novel text, by a margin worth having: 1-NN
retrieval improves from 29.2\% to 34.8\% on \textsc{ca-dgs} and from 73.9\% to 76.6\% on
\textsc{wa-amazon}. But across all four corpus-model pairings the gap narrows without
closing: 23.8 points to 19.1, 13.0 to 10.2, 22.0 to 19.6, 14.7 to 10.6.

\ifarxiv
\begin{table}[t]
\centering\small
\setlength{\tabcolsep}{4pt}
\begin{tabular}{@{}lrrrr@{}}
\toprule
Protocol & Acc & Seen & Novel & Gap \\
\midrule
\multicolumn{5}{@{}l}{\emph{\textsc{ca-dgs}}} \\
\textsc{iid-naive} & 43.8 & 54.0 & 29.6 & \textbf{24.4} \\
\textsc{iid-grouped} & 40.4 & n/a & 40.4 & n/a \\
\textsc{temporal} & 40.4 & 53.1 & 27.1 & \textbf{26.0} \\
\textsc{supplier} & 32.6 & 45.5 & 22.4 & \textbf{23.1} \\
\midrule
\multicolumn{5}{@{}l}{\emph{\textsc{wa-amazon}}} \\
\textsc{iid-naive} & 87.1 & 92.5 & 78.5 & \textbf{14.0} \\
\textsc{iid-grouped} & 83.0 & n/a & 83.0 & n/a \\
\textsc{temporal} & 84.9 & 90.6 & 82.9 & \textbf{7.7} \\
\textsc{supplier} & 85.8 & 92.9 & 80.4 & \textbf{12.5} \\
\bottomrule
\end{tabular}
\caption{Commodity-level accuracy (\%) for the fine-tuned DistilBERT, split by whether the test row's item text appeared in training. The seen-against-novel gap the classical baselines show is present here too, so a transformer with 67 million parameters does not read through the repetition to the task. \textsc{iid-grouped} has no seen rows by construction, so its Seen and Gap entries are undefined rather than zero and are marked n/a.}
\label{tab:finetune}
\end{table}
\fi

Capacity does not close it either. The fine-tuned transformer scores 54.0\% on repeated
text against 29.6\% on novel text at Commodity level on \textsc{ca-dgs} under the leaky
protocol, and holds that shape under \textsc{temporal} and \textsc{supplier}
(Table~\ref{tab:finetune}). Whatever 67 million parameters buy on this task, reading past
repetition to it is not among them.

\subsection{The Cost of Fixing the Split}

\ifarxiv
\begin{table}[t]
\centering\small
\begin{tabular}{llrrrr}
\toprule
Corpus & Model & Seg & Fam & Cls & Com \\
\midrule
\multicolumn{6}{l}{\emph{\textsc{ca-dgs}}} \\
Majority & leaky & 9.6 & 4.9 & 4.2 & 3.7 \\
 & leak-free & 9.9 & 7.0 & 4.6 & 4.1 \\
 & \emph{overstatement} & \textbf{-0.2} & \textbf{-2.1} & \textbf{-0.4} & \textbf{-0.4} \\
\addlinespace
Naive Bayes & leaky & 67.3 & 60.8 & 54.7 & 45.9 \\
 & leak-free & 63.4 & 56.9 & 50.5 & 40.8 \\
 & \emph{overstatement} & \textbf{3.9} & \textbf{3.9} & \textbf{4.2} & \textbf{5.1} \\
\addlinespace
Centroid & leaky & 53.9 & 41.5 & 30.7 & 23.3 \\
 & leak-free & 50.7 & 40.5 & 29.0 & 21.9 \\
 & \emph{overstatement} & \textbf{3.2} & \textbf{1.0} & \textbf{1.7} & \textbf{1.3} \\
\addlinespace
1-NN & leaky & 60.8 & 55.0 & 49.4 & 43.0 \\
 & leak-free & 53.8 & 47.0 & 41.7 & 34.5 \\
 & \emph{overstatement} & \textbf{7.0} & \textbf{8.1} & \textbf{7.7} & \textbf{8.6} \\
\midrule
\multicolumn{6}{l}{\emph{\textsc{wa-amazon}}} \\
Majority & leaky & 13.3 & 7.7 & 6.6 & 4.1 \\
 & leak-free & 14.1 & 7.4 & 6.5 & 4.6 \\
 & \emph{overstatement} & \textbf{-0.7} & \textbf{0.4} & \textbf{0.1} & \textbf{-0.5} \\
\addlinespace
Naive Bayes & leaky & 77.8 & 74.8 & 74.7 & 78.9 \\
 & leak-free & 77.1 & 73.6 & 72.9 & 76.4 \\
 & \emph{overstatement} & \textbf{0.8} & \textbf{1.2} & \textbf{1.8} & \textbf{2.5} \\
\addlinespace
Centroid & leaky & 64.3 & 59.5 & 61.3 & 73.0 \\
 & leak-free & 62.9 & 58.1 & 59.0 & 69.8 \\
 & \emph{overstatement} & \textbf{1.5} & \textbf{1.4} & \textbf{2.2} & \textbf{3.2} \\
\addlinespace
1-NN & leaky & 81.0 & 78.6 & 80.3 & 87.5 \\
 & leak-free & 75.4 & 72.4 & 73.4 & 79.5 \\
 & \emph{overstatement} & \textbf{5.6} & \textbf{6.2} & \textbf{6.8} & \textbf{8.0} \\
\bottomrule
\end{tabular}
\caption{Accuracy (\%) under a row-level random split (leaky) and a split over text groups (leak-free). \emph{Overstatement} is leaky minus leak-free; it is negative where the leak-free split happens to be the easier one. Wilson 95\% half-widths are below 0.4 points throughout (Appendix~\ref{app:grid}), so every learned-model gap shown far exceeds sampling error. The majority-floor rows are the exception and are not meant to clear it: a model that ignores text has no leak to gain from, and those moves are the class distribution shifting between partitions.}
\label{tab:gap}
\end{table}
\fi

Table~\ref{tab:gap} contrasts the leaking and leak-free protocols directly. The
overstatement is real, positive in all 24 lexical learned-model cells, and moderate: 0.8 to
8.6 points. The embedding centroid alone sometimes gains from grouping instead, never by a
full point.

The overstatement is also smaller than the seen-against-novel gap of the previous section. Grouping removes
identical items from the test side but not near-duplicates, shared tokens or shared brand
names, and it leaves the training set more varied, which helps. The two effects partly
cancel, so a leak-free test set is harder in one specific way rather than uniformly harder.

The ordering across models is the useful diagnostic. 1-NN retrieval loses the most, 8.6
points at Commodity on \textsc{ca-dgs}. That is exactly what should happen, because on a
leaking split retrieval simply returns the duplicate. Naive Bayes loses 5.1, the centroid
baseline 1.3. The majority-class floor barely moves, and in six of eight cells it moves the
wrong way, which is the sanity check working: a model that ignores text cannot gain from
seeing text twice, and the small negative values are the class distribution shifting between
partitions rather than any leakage effect.
The fine-tuned transformer loses 3.4 here and 4.1 on \textsc{wa-amazon},
against 1-NN's 8.6 and 8.0, so how much a leak flatters a model is a property of the
model and there is no single inflation figure to subtract from a published result. The
protocol is what needs correcting, not the number.

\subsection{Protocol Choice Against Model Choice}

\ifarxiv
\begin{table}[t]
\centering\small
\setlength{\tabcolsep}{3pt}%
\begin{tabular}{@{}llrrrr@{}}
\toprule
Corpus & Model & IID & Grouped & Temporal & Supplier \\
\midrule
\textsc{ca-dgs} & Majority & 3.7 & 4.1 & 5.2 & 0.1 \\
 & Naive Bayes & 45.9 & 40.8 & 41.0 & 33.7 \\
 & Centroid & 23.3 & 21.9 & 18.0 & 16.9 \\
 & 1-NN & 43.0 & 34.5 & 35.2 & 28.8 \\
 & DistilBERT & 43.8 & 40.4 & 40.4 & 32.6 \\
\midrule
\textsc{wa-amazon} & Majority & 4.1 & 4.6 & 2.1 & 2.5 \\
 & Naive Bayes & 78.9 & 76.4 & 78.5 & 77.8 \\
 & Centroid & 73.0 & 69.8 & 69.8 & 73.7 \\
 & 1-NN & 87.5 & 79.5 & 81.3 & 83.5 \\
 & DistilBERT & 87.1 & 83.0 & 84.9 & 85.8 \\
\bottomrule
\end{tabular}
\caption{Commodity-level accuracy (\%) across all four protocols. Column headings abbreviate \textsc{iid-naive}, \textsc{iid-grouped}, \textsc{temporal} and \textsc{supplier}, in that order. Spread along a row is the effect of evaluation design; spread down a column is the effect of model choice. The \textsc{distilbert} row is the fine-tuned transformer, on the same splits and the same 200,000-row training cap as every other model. Reading down the \textsc{wa-amazon} block, the leaky column and the leak-free columns disagree about which model is best.}
\label{tab:protocols}
\end{table}
\fi

All four protocols sit side by side at Commodity level in Table~\ref{tab:protocols}, and on
\textsc{wa-amazon} the leaky one cannot tell the two strongest models apart. Under
\textsc{iid-naive}, 1-NN retrieval reaches 87.5\% and the fine-tuned transformer 87.1\%: a
0.3-point difference, nominally to retrieval, and inside the 95\% interval on the
difference.
\ifarxiv
Each margin quoted here is the difference of the unrounded accuracies, rounded once, so it
need not equal the difference of the two rounded entries in the table.
\fi
Read that column alone and a nearest-neighbour lookup is as good as a
transformer, so the GPU is not worth buying. All three leak-free protocols separate them
in the other direction, by 3.6, 3.5 and 2.3 points to the transformer, each clearing the
interval on the difference by a factor of at least six. A leaking protocol therefore does
more than overstate a number. It can hide the comparison the evaluation was run to settle.

On \textsc{ca-dgs} the top rank holds, because naive Bayes leads under all four protocols,
and the effect moves into the margin instead: the transformer is 0.7 points ahead of 1-NN
under \textsc{iid-naive} and 5.9 ahead under \textsc{iid-grouped}. So the reversal of order
is one corpus of two and the margin change is both. Protocol choice moves the reported
number by about as much as the difference between two reasonable models, and unlike model
choice it is rarely stated in enough detail to reconstruct.

\section{What This Changed in the Product}
\label{sec:deployment}

This work began as an attempt to justify a confidence threshold. The harness ended up
changing the architecture it was built to measure, and what we thought the threshold was
for.

\paragraph{The ceiling predicted a wall that did not bind.} Section~\ref{sec:data} bounds
any classifier reading only item text at 79.4\% on terse ERP text, and
Section~\ref{sec:results} puts genuinely novel text near 32\%. As a forecast for the product
those numbers are bleak, and at the level of the product they did not materialise; where the
model actually runs, they did, almost exactly. Coverage across the three pilot
organisations rose from roughly half of lines as received to between 98.3\% and 99.3\%, and
most of that gain came from joining to evidence those organisations already owned (lines
they had coded previously, and a parts master) rather than from inference over item text.

Repetition is a measurement hazard when it straddles
a partition and an asset when it lets a deployed system recover a code its customer has
already assigned. The engineering conclusion from a low text-only ceiling is therefore not
that a better text model is needed. The problem should stop being posed as text-only. That
is what the harness showed, and a benchmark built from text and labels alone cannot say
it, which is why this section is a necessary complement.

\paragraph{Abstention became a policy rather than a defect.} Roughly 12,000 of the 920,927
lines are deliberately left uncoded. Against a coverage target that is a failure; against
the conflict ceiling it is the correct action, because for those lines the available
evidence does not determine a code, and a guess would enter the spend cube
indistinguishable from a fact. What the ceiling analysis changed was its status, from
something to apologise for into something to state.

\paragraph{The benchmark predicted the deployment, on the population it was about.}
Classification runs as a deterministic cascade: exact matches against evidence the organisation
already owns are resolved first, and the model is consulted last, on the residue no earlier
stage could settle. That residue is by construction close to the novel-text population of
Section~\ref{sec:results}: an item reaches the model precisely when no previously coded
instance of it exists. During August, reviewers acted on 1,084 such suggestions across
the three organisations and accepted 351 of them unchanged, an acceptance rate of
\textbf{32.4\%} with a Wilson 95\% interval of 29.7 to 35.2.

The benchmark's estimates for that population, taken from the corpus whose terse ERP text
resembles this data, are 31.9\% for the strongest classical baseline and 34.8\% for the
sentence-embedding retriever. Both fall inside that interval, and the observed value lies
between them. The contrast is as informative as the agreement: the same two figures on
\textsc{wa-amazon}, whose verbose marketplace titles resemble nothing in an ERP system, are
69.4\% and 76.6\%. The benchmark discriminates in the direction it should.

We are deliberately careful about what this supports. The taxonomies differ; the notion of
correctness differs, being a reviewer's live judgement rather than a buyer's recorded code;
the window is one month; and the reviewed set is the cascade's residue \emph{as sampled for
review}, not a random draw from it. The figure is also pooled, to the scope the three
organisations agreed to, and their individual rates span about nine points, one of them
below both benchmark estimates and another above both. Pooled agreement is therefore not
evidence of agreement at any single site. This is corroboration and not validation, and
$n=1{,}084$ cannot carry more than that. The conflict of interest disclosed below biases
upward, and an
interested party would have produced a number above both benchmark estimates rather than
between them. What it establishes is narrow: an accuracy estimate derived entirely from
public data landed within three points of what human reviewers did to a commercial system's
suggestions on the population that estimate was about. That the check was possible at all is
an accident of instrumentation, since a classifier that overwrites its own suggestion when
corrected keeps no evidence of how often it was wrong, and most do.

\section{Limitations}

Both corpora are US public-sector, so generalisation to private-sector or non-US spend is
untested; the two differ enough from each other to make us cautious rather than confident.
Labels are buyer-assigned rather than adjudicated by a third party, which is part of why
the conflict ceiling in Section~\ref{sec:data} exists, and it means our ground truth is
operational rather than gold. Our highest-capacity baseline is a fine-tuned DistilBERT, and
we evaluate no large language model; doing so across 1.26 million lines is a cost question,
which \citet{kapoor2024} argue is the one that accuracy-only reporting hides.
The one baseline we added to test capacity moved the finding in a direction we did not
predict: the transformer's inflation under a leaking split is smaller than
1-NN's, not larger, so leakage magnitude is model-specific and no result of ours licenses a
correction factor applied to somebody else's number. It is also one training run per cell,
so its margins carry sampling error but no seed variance, and the protocol-dependent
ordering in Section~\ref{sec:results} would be firmer with several seeds. Pilot data underlying
Section~\ref{sec:deployment} is proprietary, cannot be released, and was checked by no
independent party, so that section is self-reported and the reproducible contribution is
the public harness.

\section{Conclusion}

Accuracy claims for spend classification cannot be compared, and the reason is not
carelessness. The field has no shared measuring instrument. We built one from data that
has been public for a decade, and the first thing it showed is that the
protocol everyone uses turns a classification benchmark into a memory test for three test
rows in five. Two further properties are invisible under current reporting practice: an
annotation ceiling touching nearly half of one corpus, which bounds achievable accuracy,
and an amount field concentrated onto 0.7\% of codes, which bounds what a spend-weighted
accuracy figure can mean.

For a practitioner measuring a spend classifier on their own data, three things follow.
Group the split on normalised item text, or the reported number is partly a memory test and
the model selected on it may be the wrong one.
State the evaluable denominator at every level, because annotation depth is not uniform.
Compute the conflict ceiling on your own corpus before quoting accuracy against anyone
else's, because the ceiling is corpus-specific and a spend-weighted version of it will only
mean something if the amount field does. Each is one command against the released harness.

We think the useful unit of progress here is not a better model but an evaluation anyone
can rerun. On one of our two corpora the standard protocol cannot say which of two models is
better.

\section*{Reproducibility}

All corpora are public. Fetch-and-verify scripts, benchmark construction, protocol
definitions, baselines and every results file are at
\url{https://github.com/harshitg93/spend-classification-benchmark}.
Each statistic here is produced by a named script and recorded in a JSON results file, and
the build asserts hierarchy consistency, protocol coverage and leak-freedom on every run,
failing loudly rather than reporting silently reconciled numbers.

\section*{Acknowledgments}

The author is the founder of Cognistone, which develops SpendSide. The deployed system in
Sections~\ref{sec:application} and~\ref{sec:deployment} is therefore his own product, and
its evaluation here is not independent. Through the period this paper reports, the three
pilot organisations paid nothing to use the system. Cognistone has taken no outside
investment to date and has never been valued. The pilot figures come from those three organisations
and were produced by the product's own queries, and Section~\ref{sec:deployment}, the
paper's only external check, tests that product against a benchmark the same author built.
No independent party verified those figures, and with one author there is no co-author
check either. That limits Section~\ref{sec:deployment}. It does not limit the benchmark,
which uses only public government data and reproduces without SpendSide or any pilot
data. The three organisations agreed to publication of overall figures without being named,
so they are not named here and every figure from them is a pooled total or a range across
the three. No supplier, contract, price, spend figure or line of their data is reported.
A large language model assisted in drafting and organising this paper's prose. The
research, data, code and results are the author's own, and the author takes responsibility
for the whole text.

\bibliography{refs}

\clearpage
\appendix
\section{Software Environment}

Every cell in this paper was produced under macOS 26.6.2 on an Apple M2 with 24\,GB of
unified memory, using Python 3.12.13, scikit-learn 1.9.0, NumPy 2.5.2, SciPy 1.18.1 and
pandas 3.0.5. The embedding baselines add PyTorch 2.13.0 and sentence-transformers 6.0.0,
encoding and retrieving on the integrated GPU through the Metal Performance Shaders
backend. The fine-tuned baseline adds transformers 5.15.1 and tokenizers 0.22.2 and
trains on the same backend, at roughly 1.4 hours per cell.

\ifarxiv\else
\section{Tables Referenced from the Body}

IAAI-27 places no page limit on appendices, so the nine tables the body argues from are
placed here rather than inside the six-page limit: the seen-against-novel
decomposition (Table~\ref{tab:seenunseen}), the per-level denominators
(Table~\ref{tab:evaluable}), the leakage rates per protocol (Table~\ref{tab:leakage}),
the annotation-conflict ceiling
(Table~\ref{tab:ceiling}), the code-frequency stratification (Table~\ref{tab:tail}), the
leaky-against-leak-free contrast (Table~\ref{tab:gap}), the
lexical-against-embedding decomposition (Table~\ref{tab:neural}), the four protocols
side by side including the fine-tuned transformer (Table~\ref{tab:protocols}) and that
transformer's own seen-against-novel decomposition under each protocol
(Table~\ref{tab:finetune}). Every figure they carry is also stated either in
the prose that argues from it or in the complete grid that follows.

\begin{table}[t]
\centering
\small
\begin{tabular}{lrr}
\toprule
Protocol & \textsc{ca-dgs} & \textsc{wa-amazon} \\
\midrule
\textsc{iid-naive}   & 60.7\% & 61.2\% \\
\textsc{iid-grouped} & 0.0\%  & 0.0\%  \\
\textsc{temporal}    & 51.9\% & 27.2\% \\
\textsc{supplier}    & 45.6\% & 41.9\% \\
\bottomrule
\end{tabular}
\caption{Share of test rows whose normalised item text already appears in training. Under a
row-level shuffle roughly three in five test items have been seen already.}
\label{tab:leakage}
\end{table}

\begin{table}[t]
\centering
\small
\begin{tabular}{llrr}
\toprule
Corpus & Level & Evaluable rows & Labels \\
\midrule
\textsc{ca-dgs} & Segment   & 343,484 (100.0\%) & 56 \\
                & Family    & 343,484 (100.0\%) & 411 \\
                & Class     & 343,484 (100.0\%) & 2,377 \\
                & Commodity & 343,467 (99.995\%) & 13,399 \\
\midrule
\textsc{wa-amazon} & Segment   & 914,683 (100.0\%) & 51 \\
                   & Family    & 870,254 (95.1\%)  & 273 \\
                   & Class     & 757,434 (82.8\%)  & 996 \\
                   & Commodity & 365,085 (39.9\%)  & 2,629 \\
\bottomrule
\end{tabular}
\caption{Rows evaluable at each hierarchy level. Depth is heterogeneous, so the
denominator changes with level and must be stated. The \textsc{ca-dgs} Commodity share is
99.995\% rather than 100\%: 17 rows carry a six-digit code and are evaluable no deeper
than Class. Rounding it to 100.0\% would have retired those rows by presentation, which is
the failure this column exists to prevent.}
\label{tab:evaluable}
\end{table}

\begin{table}[t]
\centering
\small
\setlength{\tabcolsep}{4pt}
\begin{tabular}{llrrr}
\toprule
Corpus & Level & \% rows in & \multicolumn{2}{c}{Ceiling (\%)} \\
\cmidrule(l){4-5}
       &       & conflict   & count-wtd & spend-wtd \\
\midrule
\textsc{ca-dgs} & Segment   & 40.0 & 89.2 & 98.5 \\
                & Family    & 42.6 & 86.0 & 98.0 \\
                & Class     & 44.6 & 83.5 & 97.8 \\
                & Commodity & 47.6 & \textbf{79.4} & \textbf{97.4} \\
\midrule
\textsc{wa-amazon} & Segment   & 3.5 & 99.2 & 99.2 \\
                   & Family    & 4.2 & 99.1 & 99.0 \\
                   & Class     & 4.7 & 99.0 & 98.8 \\
                   & Commodity & 6.9 & \textbf{98.4} & \textbf{98.4} \\
\bottomrule
\end{tabular}
\caption{Highest accuracy achievable by any classifier that sees only item text. Computed on rows annotated to full commodity depth so that the denominator is constant across levels (343,467 rows for \textsc{ca-dgs}, 365,085 for \textsc{wa-amazon}); a coarse label is a prefix of the same row's fine label, so conflict rises and the ceiling falls with depth, as observed.}
\label{tab:ceiling}
\end{table}

\begin{table}[t]
\centering
\small
\begin{tabular}{lrrrr}
\toprule
Code frequency & Codes & \% codes & \% rows & \% spend \\
\midrule
$<5$        & 7,293 & 54.4 & 4.0  & 1.8  \\
5--49       & 4,994 & 37.3 & 22.3 & 9.9  \\
50--499     & 1,023 & 7.6  & 39.0 & 12.4 \\
$\geq 500$  & 89    & 0.7  & 34.7 & 76.0 \\
\bottomrule
\end{tabular}
\caption{\textsc{ca-dgs} commodity codes by frequency. Over half of all codes have fewer
than five examples and jointly account for 1.8\% of spend; 89 codes carry 76.0\%. Each
figure is rounded from the exact value rather than from its neighbours, so the spend
column reads 100.1\%; the underlying shares sum to 100.}
\label{tab:tail}
\end{table}

\begin{table}[t]
\centering\small
\begin{tabular}{llrrr}
\toprule
Corpus & Model & Seen & Novel & Gap \\
\midrule
\textsc{ca-dgs} & Naive Bayes & 56.0 & 31.9 & \textbf{24.1} \\
 & Centroid & 28.7 & 15.7 & \textbf{13.0} \\
 & 1-NN & 53.0 & 29.2 & \textbf{23.8} \\
\midrule
\textsc{wa-amazon} & Naive Bayes & 84.9 & 69.4 & \textbf{15.5} \\
 & Centroid & 78.6 & 63.9 & \textbf{14.7} \\
 & 1-NN & 95.9 & 73.9 & \textbf{22.0} \\
\bottomrule
\end{tabular}
\caption{Commodity-level accuracy (\%) under the leaky protocol, separated by whether the test row's item text appeared in training. A single headline number averages these two populations and reports neither.}
\label{tab:seenunseen}
\end{table}

\begin{table}[t]
\centering\small
\begin{tabular}{llrrr}
\toprule
Corpus & Representation & Seen & Novel & Gap \\
\midrule
\textsc{ca-dgs} & 1-NN & 53.0 & 29.2 & \textbf{23.8} \\
 & 1-NN (embed.) & 53.9 & 34.8 & \textbf{19.1} \\
\addlinespace
 & Centroid & 28.7 & 15.7 & \textbf{13.0} \\
 & Centroid (embed.) & 32.2 & 22.0 & \textbf{10.2} \\
\addlinespace
\midrule
\textsc{wa-amazon} & 1-NN & 95.9 & 73.9 & \textbf{22.0} \\
 & 1-NN (embed.) & 96.3 & 76.6 & \textbf{19.6} \\
\addlinespace
 & Centroid & 78.6 & 63.9 & \textbf{14.7} \\
 & Centroid (embed.) & 77.3 & 66.7 & \textbf{10.6} \\
\addlinespace
\bottomrule
\end{tabular}
\caption{Commodity-level accuracy (\%) under the leaky protocol, lexical TF-IDF against \texttt{all-MiniLM-L6-v2} sentence embeddings, on identical splits with identical scoring. The seen-against-novel gap does not close under a semantic representation, so it is a property of the task rather than of lexical matching.}
\label{tab:neural}
\end{table}

\begin{table}[t]
\centering\small
\setlength{\tabcolsep}{3pt}%
\begin{tabular}{@{}llrrrr@{}}
\toprule
Corpus & Model & IID & Grouped & Temporal & Supplier \\
\midrule
\textsc{ca-dgs} & Majority & 3.7 & 4.1 & 5.2 & 0.1 \\
 & Naive Bayes & 45.9 & 40.8 & 41.0 & 33.7 \\
 & Centroid & 23.3 & 21.9 & 18.0 & 16.9 \\
 & 1-NN & 43.0 & 34.5 & 35.2 & 28.8 \\
 & DistilBERT & 43.8 & 40.4 & 40.4 & 32.6 \\
\midrule
\textsc{wa-amazon} & Majority & 4.1 & 4.6 & 2.1 & 2.5 \\
 & Naive Bayes & 78.9 & 76.4 & 78.5 & 77.8 \\
 & Centroid & 73.0 & 69.8 & 69.8 & 73.7 \\
 & 1-NN & 87.5 & 79.5 & 81.3 & 83.5 \\
 & DistilBERT & 87.1 & 83.0 & 84.9 & 85.8 \\
\bottomrule
\end{tabular}
\caption{Commodity-level accuracy (\%) across all four protocols. Column headings abbreviate \textsc{iid-naive}, \textsc{iid-grouped}, \textsc{temporal} and \textsc{supplier}, in that order. Spread along a row is the effect of evaluation design; spread down a column is the effect of model choice. The \textsc{distilbert} row is the fine-tuned transformer, on the same splits and the same 200,000-row training cap as every other model. Reading down the \textsc{wa-amazon} block, the leaky column and the leak-free columns disagree about which model is best.}
\label{tab:protocols}
\end{table}

\begin{table}[t]
\centering\small
\setlength{\tabcolsep}{4pt}
\begin{tabular}{@{}lrrrr@{}}
\toprule
Protocol & Acc & Seen & Novel & Gap \\
\midrule
\multicolumn{5}{@{}l}{\emph{\textsc{ca-dgs}}} \\
\textsc{iid-naive} & 43.8 & 54.0 & 29.6 & \textbf{24.4} \\
\textsc{iid-grouped} & 40.4 & n/a & 40.4 & n/a \\
\textsc{temporal} & 40.4 & 53.1 & 27.1 & \textbf{26.0} \\
\textsc{supplier} & 32.6 & 45.5 & 22.4 & \textbf{23.1} \\
\midrule
\multicolumn{5}{@{}l}{\emph{\textsc{wa-amazon}}} \\
\textsc{iid-naive} & 87.1 & 92.5 & 78.5 & \textbf{14.0} \\
\textsc{iid-grouped} & 83.0 & n/a & 83.0 & n/a \\
\textsc{temporal} & 84.9 & 90.6 & 82.9 & \textbf{7.7} \\
\textsc{supplier} & 85.8 & 92.9 & 80.4 & \textbf{12.5} \\
\bottomrule
\end{tabular}
\caption{Commodity-level accuracy (\%) for the fine-tuned DistilBERT, split by whether the test row's item text appeared in training. The seen-against-novel gap the classical baselines show is present here too, so a transformer with 67 million parameters does not read through the repetition to the task. \textsc{iid-grouped} has no seen rows by construction, so its Seen and Gap entries are undefined rather than zero and are marked n/a.}
\label{tab:finetune}
\end{table}

\fi

\section{Macro and Micro Averages Are Different Metrics}

\ifarxiv
\begin{table}[t]
\centering
\small
\begin{tabular}{lrrrr}
\toprule
Code frequency & Codes & \% codes & \% rows & \% spend \\
\midrule
$<5$        & 7,293 & 54.4 & 4.0  & 1.8  \\
5--49       & 4,994 & 37.3 & 22.3 & 9.9  \\
50--499     & 1,023 & 7.6  & 39.0 & 12.4 \\
$\geq 500$  & 89    & 0.7  & 34.7 & 76.0 \\
\bottomrule
\end{tabular}
\caption{\textsc{ca-dgs} commodity codes by frequency. Over half of all codes have fewer
than five examples and jointly account for 1.8\% of spend; 89 codes carry 76.0\%. Each
figure is rounded from the exact value rather than from its neighbours, so the spend
column reads 100.1\%; the underlying shares sum to 100.}
\label{tab:tail}
\end{table}
\fi

\ifarxiv
Supporting material for Section~\ref{sec:data}.
\else
Moved here from Section~\ref{sec:data} for length.
\fi
Table~\ref{tab:tail} stratifies \textsc{ca-dgs} commodity codes by
frequency. Macro-averaged metrics are dominated by codes that barely matter financially, and
micro-averaged metrics are dominated by codes a category manager
already watches by hand. Reporting either alone is misleading, and the field reports
micro-accuracy at a single self-chosen level. The common assertion that value hides in the
long tail is not supported by this corpus's amount field, where the tail is numerous and
cheap, though for the reasons given in Section~\ref{sec:data} that is a fact about
California contracting rather than
about procurement generally.
\textsc{wa-amazon} has a head too, but a far broader one: there 151 codes (5.7\%) carry
75.3\% of spend and 76.2\% of rows, so spend tracks volume.
The case for tail analytics rests instead on the head already receiving
dedicated sourcing attention \citep{verizon2020}: the claim is about where effort is
scarce, not about where the money is.

The size of the macro-averaged figures in the complete grid follows from the same
stratification. Macro-$F_1$ weights all 11,509 \textsc{ca-dgs} commodity classes present in
training equally, most of which have a handful of examples, so at that level it comes out
well under half of each model's accuracy. On \textsc{wa-amazon}, with a label space roughly
five times narrower, every one of the same models scores higher on it. Neither number
should be read as a corpus-independent quality score.

\section{Representation Ablation}

\ifarxiv
Supporting material for Section~\ref{sec:lexical}.
\else
Moved here from Section~\ref{sec:lexical} for length.
\fi
The appendix grid adds a character
2--5-gram model at two vocabulary sizes and an uncapped word vocabulary. On
\textsc{ca-dgs}, 1-NN novel-text accuracy is 29.2\% on capped word features, 30.4\%
uncapped, 33.0\% on character n-grams and 34.8\% on embeddings, for gaps of 23.8, 23.2,
21.1 and 19.1 points. Across word, character and embedding features the gap narrows
monotonically without closing, and lifting the 30,000-feature cap adds 1.2 points on novel
text, so the cap is not what produces it.

\section{Complete Results Grid}
\label{app:grid}

The complete grid follows, as one table per corpus and protocol. Every row carries its
Wilson 95\% interval half-width and its test-set size. The baselines of
Section~\ref{sec:results} were fitted in every cell of corpus $\times$ protocol $\times$
level, so no entry is absent for them; rows for the additional representation
configurations and for the fine-tuned transformer appear only in the cells those
configurations were run on. Every figure the body quotes for the transformer is at
Commodity level.

\begin{table*}[t]
\centering\footnotesize
\begin{tabular}{lllrrrrrr}
\toprule
Corpus & Protocol & Model & Level & Acc & $\pm$95\% & Macro-$F_1$ & Spend-wtd & $n_\text{test}$ \\
\midrule
ca-dgs & iid-naive & Majority & segment & 9.6 & 0.22 & 0.3 & 0.7 & 68,757 \\
ca-dgs & iid-naive & Naive Bayes & segment & 67.3 & 0.35 & 54.6 & 86.5 & 68,757 \\
ca-dgs & iid-naive & Centroid & segment & 53.9 & 0.37 & 43.8 & 64.2 & 68,757 \\
ca-dgs & iid-naive & 1-NN & segment & 60.8 & 0.36 & 51.1 & 86.9 & 68,757 \\
ca-dgs & iid-naive & 1-NN (embed.) & segment & 64.1 & 0.36 & 54.6 & 87.1 & 68,757 \\
ca-dgs & iid-naive & Centroid (embed.) & segment & 45.9 & 0.37 & 35.0 & 83.0 & 68,757 \\
ca-dgs & iid-naive & Majority & family & 4.9 & 0.16 & 0.0 & 0.0 & 68,757 \\
ca-dgs & iid-naive & Naive Bayes & family & 60.8 & 0.36 & 33.4 & 83.2 & 68,757 \\
ca-dgs & iid-naive & Centroid & family & 41.5 & 0.37 & 25.4 & 43.4 & 68,757 \\
ca-dgs & iid-naive & 1-NN & family & 55.0 & 0.37 & 33.3 & 86.4 & 68,757 \\
ca-dgs & iid-naive & 1-NN (embed.) & family & 58.2 & 0.37 & 35.8 & 86.5 & 68,757 \\
ca-dgs & iid-naive & Centroid (embed.) & family & 39.6 & 0.37 & 22.4 & 48.1 & 68,757 \\
ca-dgs & iid-naive & Majority & class & 4.2 & 0.15 & 0.0 & 0.1 & 68,757 \\
ca-dgs & iid-naive & Naive Bayes & class & 54.7 & 0.37 & 20.4 & 82.2 & 68,757 \\
ca-dgs & iid-naive & Centroid & class & 30.7 & 0.34 & 16.5 & 35.9 & 68,757 \\
ca-dgs & iid-naive & 1-NN & class & 49.4 & 0.37 & 22.5 & 85.8 & 68,757 \\
ca-dgs & iid-naive & 1-NN (embed.) & class & 52.4 & 0.37 & 25.1 & 85.7 & 68,757 \\
ca-dgs & iid-naive & Centroid (embed.) & class & 33.4 & 0.35 & 16.0 & 29.5 & 68,757 \\
ca-dgs & iid-naive & Majority & commodity & 3.7 & 0.14 & 0.0 & 0.0 & 68,752 \\
ca-dgs & iid-naive & Naive Bayes & commodity & 45.9 & 0.37 & 11.3 & 82.2 & 68,752 \\
ca-dgs & iid-naive & Centroid & commodity & 23.3 & 0.32 & 10.8 & 16.9 & 68,752 \\
ca-dgs & iid-naive & 1-NN & commodity & 43.0 & 0.37 & 14.7 & 84.8 & 68,752 \\
ca-dgs & iid-naive & 1-NN (embed.) & commodity & 45.9 & 0.37 & 16.8 & 86.2 & 68,752 \\
ca-dgs & iid-naive & Centroid (embed.) & commodity & 28.0 & 0.34 & 12.4 & 16.6 & 68,752 \\
ca-dgs & iid-naive & Centroid (char 2--5) & commodity & 25.4 & 0.33 & 12.3 & 48.6 & 68,752 \\
ca-dgs & iid-naive & 1-NN (char 2--5) & commodity & 45.2 & 0.37 & 16.5 & 86.2 & 68,752 \\
ca-dgs & iid-naive & Naive Bayes (char 2--5, 30k) & commodity & 38.0 & 0.36 & 5.0 & 78.1 & 68,752 \\
ca-dgs & iid-naive & Centroid (char 2--5, 30k) & commodity & 24.4 & 0.32 & 11.6 & 24.1 & 68,752 \\
ca-dgs & iid-naive & 1-NN (char 2--5, 30k) & commodity & 45.2 & 0.37 & 16.4 & 86.9 & 68,752 \\
ca-dgs & iid-naive & Centroid (word 1--2, uncapped) & commodity & 27.1 & 0.33 & 12.9 & 18.3 & 68,752 \\
ca-dgs & iid-naive & 1-NN (word 1--2, uncapped) & commodity & 43.9 & 0.37 & 15.6 & 85.4 & 68,752 \\
ca-dgs & iid-naive & DistilBERT (fine-tuned) & commodity & 43.8 & 0.37 & 5.1 & 84.6 & 68,752 \\
\bottomrule
\end{tabular}
\caption{Complete results grid, part 1 of 8: \textsc{ca-dgs} under \textsc{iid-naive}. $\pm$95\% is the Wilson score interval half-width in percentage points. Because item text repeats, rows are not fully independent on the leaking protocols, so these intervals are a lower bound on true uncertainty. The remaining corpus-protocol blocks follow in Tables~\ref{tab:full-ca-dgs-iid-grouped}--\ref{tab:full-wa-amazon-supplier}.}
\label{tab:full}
\end{table*}

\begin{table*}[t]
\centering\footnotesize
\begin{tabular}{lllrrrrrr}
\toprule
Corpus & Protocol & Model & Level & Acc & $\pm$95\% & Macro-$F_1$ & Spend-wtd & $n_\text{test}$ \\
\midrule
ca-dgs & iid-grouped & Majority & segment & 9.9 & 0.22 & 0.3 & 1.5 & 68,697 \\
ca-dgs & iid-grouped & Naive Bayes & segment & 63.4 & 0.36 & 47.1 & 79.5 & 68,697 \\
ca-dgs & iid-grouped & Centroid & segment & 50.7 & 0.37 & 36.9 & 76.1 & 68,697 \\
ca-dgs & iid-grouped & 1-NN & segment & 53.8 & 0.37 & 39.8 & 72.0 & 68,697 \\
ca-dgs & iid-grouped & 1-NN (embed.) & segment & 63.2 & 0.36 & 49.1 & 80.6 & 68,697 \\
ca-dgs & iid-grouped & Centroid (embed.) & segment & 46.5 & 0.37 & 31.6 & 76.0 & 68,697 \\
ca-dgs & iid-grouped & Majority & family & 7.0 & 0.19 & 0.0 & 0.1 & 68,697 \\
ca-dgs & iid-grouped & Naive Bayes & family & 56.9 & 0.37 & 28.3 & 76.7 & 68,697 \\
ca-dgs & iid-grouped & Centroid & family & 40.5 & 0.37 & 21.1 & 56.6 & 68,697 \\
ca-dgs & iid-grouped & 1-NN & family & 47.0 & 0.37 & 25.1 & 67.9 & 68,697 \\
ca-dgs & iid-grouped & 1-NN (embed.) & family & 56.3 & 0.37 & 31.5 & 78.2 & 68,697 \\
ca-dgs & iid-grouped & Centroid (embed.) & family & 40.0 & 0.37 & 20.0 & 62.8 & 68,697 \\
ca-dgs & iid-grouped & Majority & class & 4.6 & 0.16 & 0.0 & 0.1 & 68,697 \\
ca-dgs & iid-grouped & Naive Bayes & class & 50.5 & 0.37 & 16.5 & 76.8 & 68,697 \\
ca-dgs & iid-grouped & Centroid & class & 29.0 & 0.34 & 12.8 & 21.7 & 68,697 \\
ca-dgs & iid-grouped & 1-NN & class & 41.7 & 0.37 & 16.0 & 67.4 & 68,697 \\
ca-dgs & iid-grouped & 1-NN (embed.) & class & 48.9 & 0.37 & 20.2 & 77.1 & 68,697 \\
ca-dgs & iid-grouped & Centroid (embed.) & class & 28.9 & 0.34 & 13.4 & 29.8 & 68,697 \\
ca-dgs & iid-grouped & Majority & commodity & 4.1 & 0.15 & 0.0 & 0.1 & 68,694 \\
ca-dgs & iid-grouped & Naive Bayes & commodity & 40.8 & 0.37 & 8.6 & 74.7 & 68,694 \\
ca-dgs & iid-grouped & Centroid & commodity & 21.9 & 0.31 & 8.1 & 4.6 & 68,694 \\
ca-dgs & iid-grouped & 1-NN & commodity & 34.5 & 0.36 & 10.4 & 67.3 & 68,694 \\
ca-dgs & iid-grouped & 1-NN (embed.) & commodity & 39.7 & 0.37 & 13.1 & 72.2 & 68,694 \\
ca-dgs & iid-grouped & Centroid (embed.) & commodity & 25.1 & 0.32 & 10.0 & 19.2 & 68,694 \\
ca-dgs & iid-grouped & DistilBERT (fine-tuned) & commodity & 40.4 & 0.37 & 4.0 & 75.3 & 68,694 \\
\bottomrule
\end{tabular}
\caption{Complete results grid, part 2 of 8: \textsc{ca-dgs} under \textsc{iid-grouped}. Columns as in Table~\ref{tab:full}.}
\label{tab:full-ca-dgs-iid-grouped}
\end{table*}

\begin{table*}[t]
\centering\footnotesize
\begin{tabular}{lllrrrrrr}
\toprule
Corpus & Protocol & Model & Level & Acc & $\pm$95\% & Macro-$F_1$ & Spend-wtd & $n_\text{test}$ \\
\midrule
ca-dgs & temporal & Majority & segment & 9.5 & 0.17 & 0.3 & 1.5 & 115,741 \\
ca-dgs & temporal & Naive Bayes & segment & 64.2 & 0.28 & 50.6 & 67.9 & 115,741 \\
ca-dgs & temporal & Centroid & segment & 51.5 & 0.29 & 41.5 & 63.2 & 115,741 \\
ca-dgs & temporal & 1-NN & segment & 55.9 & 0.29 & 44.1 & 57.6 & 115,741 \\
ca-dgs & temporal & 1-NN (embed.) & segment & 58.9 & 0.28 & 47.4 & 70.7 & 115,741 \\
ca-dgs & temporal & Centroid (embed.) & segment & 46.7 & 0.29 & 35.2 & 65.6 & 115,741 \\
ca-dgs & temporal & Majority & family & 4.4 & 0.12 & 0.0 & 0.1 & 115,741 \\
ca-dgs & temporal & Naive Bayes & family & 57.4 & 0.28 & 28.0 & 65.6 & 115,741 \\
ca-dgs & temporal & Centroid & family & 35.2 & 0.28 & 23.0 & 32.9 & 115,741 \\
ca-dgs & temporal & 1-NN & family & 49.5 & 0.29 & 25.1 & 55.0 & 115,741 \\
ca-dgs & temporal & 1-NN (embed.) & family & 52.4 & 0.29 & 27.9 & 69.0 & 115,741 \\
ca-dgs & temporal & Centroid (embed.) & family & 40.0 & 0.28 & 21.6 & 46.2 & 115,741 \\
ca-dgs & temporal & Majority & class & 5.7 & 0.13 & 0.0 & 0.1 & 115,741 \\
ca-dgs & temporal & Naive Bayes & class & 50.9 & 0.29 & 15.3 & 64.6 & 115,741 \\
ca-dgs & temporal & Centroid & class & 25.9 & 0.25 & 12.7 & 23.3 & 115,741 \\
ca-dgs & temporal & 1-NN & class & 43.6 & 0.29 & 14.6 & 54.0 & 115,741 \\
ca-dgs & temporal & 1-NN (embed.) & class & 46.3 & 0.29 & 16.3 & 67.6 & 115,741 \\
ca-dgs & temporal & Centroid (embed.) & class & 29.1 & 0.26 & 12.7 & 43.6 & 115,741 \\
ca-dgs & temporal & Majority & commodity & 5.2 & 0.13 & 0.0 & 0.0 & 115,741 \\
ca-dgs & temporal & Naive Bayes & commodity & 41.0 & 0.28 & 7.2 & 63.2 & 115,741 \\
ca-dgs & temporal & Centroid & commodity & 18.0 & 0.22 & 6.9 & 20.5 & 115,741 \\
ca-dgs & temporal & 1-NN & commodity & 35.2 & 0.28 & 8.0 & 58.7 & 115,741 \\
ca-dgs & temporal & 1-NN (embed.) & commodity & 37.4 & 0.28 & 9.2 & 65.4 & 115,741 \\
ca-dgs & temporal & Centroid (embed.) & commodity & 24.1 & 0.25 & 8.0 & 17.2 & 115,741 \\
ca-dgs & temporal & DistilBERT (fine-tuned) & commodity & 40.4 & 0.28 & 3.8 & 66.6 & 115,741 \\
\bottomrule
\end{tabular}
\caption{Complete results grid, part 3 of 8: \textsc{ca-dgs} under \textsc{temporal}. Columns as in Table~\ref{tab:full}.}
\label{tab:full-ca-dgs-temporal}
\end{table*}

\begin{table*}[t]
\centering\footnotesize
\begin{tabular}{lllrrrrrr}
\toprule
Corpus & Protocol & Model & Level & Acc & $\pm$95\% & Macro-$F_1$ & Spend-wtd & $n_\text{test}$ \\
\midrule
ca-dgs & supplier & Majority & segment & 7.4 & 0.20 & 0.2 & 2.1 & 68,699 \\
ca-dgs & supplier & Naive Bayes & segment & 57.9 & 0.37 & 46.5 & 62.9 & 68,699 \\
ca-dgs & supplier & Centroid & segment & 46.8 & 0.37 & 37.8 & 59.8 & 68,699 \\
ca-dgs & supplier & 1-NN & segment & 51.6 & 0.37 & 42.1 & 62.2 & 68,699 \\
ca-dgs & supplier & 1-NN (embed.) & segment & 55.5 & 0.37 & 45.7 & 66.6 & 68,699 \\
ca-dgs & supplier & Centroid (embed.) & segment & 41.7 & 0.37 & 30.8 & 58.0 & 68,699 \\
ca-dgs & supplier & Majority & family & 2.0 & 0.10 & 0.0 & 0.1 & 68,699 \\
ca-dgs & supplier & Naive Bayes & family & 50.3 & 0.37 & 26.3 & 61.5 & 68,699 \\
ca-dgs & supplier & Centroid & family & 33.1 & 0.35 & 20.9 & 42.5 & 68,699 \\
ca-dgs & supplier & 1-NN & family & 44.3 & 0.37 & 24.0 & 60.7 & 68,699 \\
ca-dgs & supplier & 1-NN (embed.) & family & 48.1 & 0.37 & 27.5 & 65.2 & 68,699 \\
ca-dgs & supplier & Centroid (embed.) & family & 34.8 & 0.36 & 19.3 & 46.4 & 68,699 \\
ca-dgs & supplier & Majority & class & 0.2 & 0.03 & 0.0 & 0.0 & 68,699 \\
ca-dgs & supplier & Naive Bayes & class & 43.5 & 0.37 & 14.5 & 59.2 & 68,699 \\
ca-dgs & supplier & Centroid & class & 25.5 & 0.33 & 11.9 & 27.2 & 68,699 \\
ca-dgs & supplier & 1-NN & class & 37.8 & 0.36 & 14.3 & 57.4 & 68,699 \\
ca-dgs & supplier & 1-NN (embed.) & class & 41.5 & 0.37 & 16.5 & 62.1 & 68,699 \\
ca-dgs & supplier & Centroid (embed.) & class & 28.7 & 0.34 & 12.2 & 41.8 & 68,699 \\
ca-dgs & supplier & Majority & commodity & 0.1 & 0.02 & 0.0 & 0.0 & 68,698 \\
ca-dgs & supplier & Naive Bayes & commodity & 33.7 & 0.35 & 7.2 & 55.4 & 68,698 \\
ca-dgs & supplier & Centroid & commodity & 16.9 & 0.28 & 6.6 & 11.7 & 68,698 \\
ca-dgs & supplier & 1-NN & commodity & 28.8 & 0.34 & 7.7 & 50.6 & 68,698 \\
ca-dgs & supplier & 1-NN (embed.) & commodity & 31.7 & 0.35 & 9.2 & 57.6 & 68,698 \\
ca-dgs & supplier & Centroid (embed.) & commodity & 21.1 & 0.31 & 8.1 & 23.2 & 68,698 \\
ca-dgs & supplier & DistilBERT (fine-tuned) & commodity & 32.6 & 0.35 & 3.9 & 55.4 & 68,698 \\
\bottomrule
\end{tabular}
\caption{Complete results grid, part 4 of 8: \textsc{ca-dgs} under \textsc{supplier}. Columns as in Table~\ref{tab:full}.}
\label{tab:full-ca-dgs-supplier}
\end{table*}

\begin{table*}[t]
\centering\footnotesize
\begin{tabular}{lllrrrrrr}
\toprule
Corpus & Protocol & Model & Level & Acc & $\pm$95\% & Macro-$F_1$ & Spend-wtd & $n_\text{test}$ \\
\midrule
wa-amazon & iid-naive & Majority & segment & 13.3 & 0.16 & 0.5 & 19.7 & 182,546 \\
wa-amazon & iid-naive & Naive Bayes & segment & 77.8 & 0.19 & 41.7 & 76.3 & 182,546 \\
wa-amazon & iid-naive & Centroid & segment & 64.3 & 0.22 & 36.4 & 61.1 & 182,546 \\
wa-amazon & iid-naive & 1-NN & segment & 81.0 & 0.18 & 57.5 & 80.7 & 182,546 \\
wa-amazon & iid-naive & 1-NN (embed.) & segment & 82.7 & 0.17 & 59.8 & 82.6 & 182,546 \\
wa-amazon & iid-naive & Centroid (embed.) & segment & 57.1 & 0.23 & 31.3 & 54.3 & 182,546 \\
wa-amazon & iid-naive & Majority & family & 7.7 & 0.13 & 0.1 & 12.4 & 173,648 \\
wa-amazon & iid-naive & Naive Bayes & family & 74.8 & 0.20 & 26.7 & 74.1 & 173,648 \\
wa-amazon & iid-naive & Centroid & family & 59.5 & 0.23 & 26.0 & 55.6 & 173,648 \\
wa-amazon & iid-naive & 1-NN & family & 78.6 & 0.19 & 45.7 & 78.9 & 173,648 \\
wa-amazon & iid-naive & 1-NN (embed.) & family & 80.6 & 0.19 & 48.5 & 80.8 & 173,648 \\
wa-amazon & iid-naive & Centroid (embed.) & family & 55.3 & 0.23 & 24.6 & 51.4 & 173,648 \\
wa-amazon & iid-naive & Majority & class & 6.6 & 0.13 & 0.0 & 4.3 & 151,211 \\
wa-amazon & iid-naive & Naive Bayes & class & 74.7 & 0.22 & 21.8 & 73.9 & 151,211 \\
wa-amazon & iid-naive & Centroid & class & 61.3 & 0.25 & 25.4 & 59.9 & 151,211 \\
wa-amazon & iid-naive & 1-NN & class & 80.3 & 0.20 & 42.9 & 80.6 & 151,211 \\
wa-amazon & iid-naive & 1-NN (embed.) & class & 82.2 & 0.19 & 44.8 & 82.7 & 151,211 \\
wa-amazon & iid-naive & Centroid (embed.) & class & 59.2 & 0.25 & 26.8 & 57.2 & 151,211 \\
wa-amazon & iid-naive & Majority & commodity & 4.1 & 0.14 & 0.0 & 2.2 & 72,643 \\
wa-amazon & iid-naive & Naive Bayes & commodity & 78.9 & 0.30 & 22.8 & 78.6 & 72,643 \\
wa-amazon & iid-naive & Centroid & commodity & 73.0 & 0.32 & 34.3 & 73.5 & 72,643 \\
wa-amazon & iid-naive & 1-NN & commodity & 87.5 & 0.24 & 52.3 & 88.1 & 72,643 \\
wa-amazon & iid-naive & 1-NN (embed.) & commodity & 88.7 & 0.23 & 55.7 & 89.3 & 72,643 \\
wa-amazon & iid-naive & Centroid (embed.) & commodity & 73.2 & 0.32 & 40.6 & 74.2 & 72,643 \\
wa-amazon & iid-naive & Centroid (char 2--5) & commodity & 74.9 & 0.32 & 38.9 & 75.5 & 72,643 \\
wa-amazon & iid-naive & 1-NN (char 2--5) & commodity & 88.0 & 0.24 & 54.6 & 88.5 & 72,643 \\
wa-amazon & iid-naive & Naive Bayes (char 2--5, 30k) & commodity & 68.0 & 0.34 & 10.9 & 66.8 & 72,643 \\
wa-amazon & iid-naive & Centroid (char 2--5, 30k) & commodity & 73.2 & 0.32 & 37.7 & 73.6 & 72,643 \\
wa-amazon & iid-naive & 1-NN (char 2--5, 30k) & commodity & 88.3 & 0.23 & 55.3 & 88.5 & 72,643 \\
wa-amazon & iid-naive & Centroid (word 1--2, uncapped) & commodity & 78.0 & 0.30 & 42.5 & 78.4 & 72,643 \\
wa-amazon & iid-naive & 1-NN (word 1--2, uncapped) & commodity & 87.8 & 0.24 & 54.7 & 88.4 & 72,643 \\
wa-amazon & iid-naive & DistilBERT (fine-tuned) & commodity & 87.1 & 0.24 & 31.2 & 87.0 & 72,643 \\
\bottomrule
\end{tabular}
\caption{Complete results grid, part 5 of 8: \textsc{wa-amazon} under \textsc{iid-naive}. Columns as in Table~\ref{tab:full}.}
\label{tab:full-wa-amazon-iid-naive}
\end{table*}

\begin{table*}[t]
\centering\footnotesize
\begin{tabular}{lllrrrrrr}
\toprule
Corpus & Protocol & Model & Level & Acc & $\pm$95\% & Macro-$F_1$ & Spend-wtd & $n_\text{test}$ \\
\midrule
wa-amazon & iid-grouped & Majority & segment & 14.1 & 0.16 & 0.5 & 21.5 & 182,938 \\
wa-amazon & iid-grouped & Naive Bayes & segment & 77.1 & 0.19 & 41.3 & 75.9 & 182,938 \\
wa-amazon & iid-grouped & Centroid & segment & 62.9 & 0.22 & 34.5 & 58.6 & 182,938 \\
wa-amazon & iid-grouped & 1-NN & segment & 75.4 & 0.20 & 43.9 & 73.8 & 182,938 \\
wa-amazon & iid-grouped & 1-NN (embed.) & segment & 77.6 & 0.19 & 46.7 & 76.5 & 182,938 \\
wa-amazon & iid-grouped & Centroid (embed.) & segment & 56.9 & 0.23 & 30.7 & 53.0 & 182,938 \\
wa-amazon & iid-grouped & Majority & family & 7.4 & 0.12 & 0.1 & 11.9 & 174,129 \\
wa-amazon & iid-grouped & Naive Bayes & family & 73.6 & 0.21 & 25.7 & 72.5 & 174,129 \\
wa-amazon & iid-grouped & Centroid & family & 58.1 & 0.23 & 23.9 & 54.0 & 174,129 \\
wa-amazon & iid-grouped & 1-NN & family & 72.4 & 0.21 & 33.7 & 71.7 & 174,129 \\
wa-amazon & iid-grouped & 1-NN (embed.) & family & 74.3 & 0.21 & 35.1 & 73.8 & 174,129 \\
wa-amazon & iid-grouped & Centroid (embed.) & family & 55.4 & 0.23 & 22.9 & 51.1 & 174,129 \\
wa-amazon & iid-grouped & Majority & class & 6.5 & 0.12 & 0.0 & 4.1 & 151,562 \\
wa-amazon & iid-grouped & Naive Bayes & class & 72.9 & 0.22 & 20.4 & 71.8 & 151,562 \\
wa-amazon & iid-grouped & Centroid & class & 59.0 & 0.25 & 22.5 & 58.2 & 151,562 \\
wa-amazon & iid-grouped & 1-NN & class & 73.4 & 0.22 & 30.6 & 72.6 & 151,562 \\
wa-amazon & iid-grouped & 1-NN (embed.) & class & 75.9 & 0.22 & 33.7 & 75.0 & 151,562 \\
wa-amazon & iid-grouped & Centroid (embed.) & class & 59.4 & 0.25 & 24.4 & 57.9 & 151,562 \\
wa-amazon & iid-grouped & Majority & commodity & 4.6 & 0.15 & 0.0 & 2.8 & 72,444 \\
wa-amazon & iid-grouped & Naive Bayes & commodity & 76.4 & 0.31 & 19.8 & 75.7 & 72,444 \\
wa-amazon & iid-grouped & Centroid & commodity & 69.8 & 0.33 & 27.3 & 70.9 & 72,444 \\
wa-amazon & iid-grouped & 1-NN & commodity & 79.5 & 0.29 & 36.1 & 79.2 & 72,444 \\
wa-amazon & iid-grouped & 1-NN (embed.) & commodity & 81.2 & 0.28 & 39.7 & 81.5 & 72,444 \\
wa-amazon & iid-grouped & Centroid (embed.) & commodity & 70.7 & 0.33 & 32.6 & 72.3 & 72,444 \\
wa-amazon & iid-grouped & DistilBERT (fine-tuned) & commodity & 83.0 & 0.27 & 26.8 & 82.8 & 72,444 \\
\bottomrule
\end{tabular}
\caption{Complete results grid, part 6 of 8: \textsc{wa-amazon} under \textsc{iid-grouped}. Columns as in Table~\ref{tab:full}.}
\label{tab:full-wa-amazon-iid-grouped}
\end{table*}

\begin{table*}[t]
\centering\footnotesize
\begin{tabular}{lllrrrrrr}
\toprule
Corpus & Protocol & Model & Level & Acc & $\pm$95\% & Macro-$F_1$ & Spend-wtd & $n_\text{test}$ \\
\midrule
wa-amazon & temporal & Majority & segment & 9.3 & 0.12 & 0.4 & 16.6 & 211,001 \\
wa-amazon & temporal & Naive Bayes & segment & 80.1 & 0.17 & 49.1 & 79.5 & 211,001 \\
wa-amazon & temporal & Centroid & segment & 66.6 & 0.20 & 36.9 & 63.7 & 211,001 \\
wa-amazon & temporal & 1-NN & segment & 78.4 & 0.18 & 47.2 & 78.2 & 211,001 \\
wa-amazon & temporal & 1-NN (embed.) & segment & 80.2 & 0.17 & 49.5 & 80.9 & 211,001 \\
wa-amazon & temporal & Centroid (embed.) & segment & 61.6 & 0.21 & 33.5 & 59.1 & 211,001 \\
wa-amazon & temporal & Majority & family & 8.6 & 0.12 & 0.1 & 13.8 & 208,372 \\
wa-amazon & temporal & Naive Bayes & family & 74.9 & 0.19 & 27.0 & 73.1 & 208,372 \\
wa-amazon & temporal & Centroid & family & 60.8 & 0.21 & 24.6 & 57.2 & 208,372 \\
wa-amazon & temporal & 1-NN & family & 73.7 & 0.19 & 32.8 & 73.6 & 208,372 \\
wa-amazon & temporal & 1-NN (embed.) & family & 75.4 & 0.18 & 35.2 & 75.1 & 208,372 \\
wa-amazon & temporal & Centroid (embed.) & family & 56.7 & 0.21 & 23.5 & 52.3 & 208,372 \\
wa-amazon & temporal & Majority & class & 4.0 & 0.09 & 0.0 & 2.7 & 194,754 \\
wa-amazon & temporal & Naive Bayes & class & 70.8 & 0.20 & 21.6 & 70.8 & 194,754 \\
wa-amazon & temporal & Centroid & class & 59.8 & 0.22 & 20.7 & 59.9 & 194,754 \\
wa-amazon & temporal & 1-NN & class & 72.0 & 0.20 & 28.0 & 72.1 & 194,754 \\
wa-amazon & temporal & 1-NN (embed.) & class & 74.1 & 0.19 & 30.5 & 74.5 & 194,754 \\
wa-amazon & temporal & Centroid (embed.) & class & 58.9 & 0.22 & 22.2 & 58.4 & 194,754 \\
wa-amazon & temporal & Majority & commodity & 2.1 & 0.10 & 0.0 & 0.9 & 74,881 \\
wa-amazon & temporal & Naive Bayes & commodity & 78.5 & 0.29 & 23.8 & 78.3 & 74,881 \\
wa-amazon & temporal & Centroid & commodity & 69.8 & 0.33 & 24.5 & 71.4 & 74,881 \\
wa-amazon & temporal & 1-NN & commodity & 81.3 & 0.28 & 34.2 & 81.8 & 74,881 \\
wa-amazon & temporal & 1-NN (embed.) & commodity & 82.4 & 0.27 & 37.4 & 82.9 & 74,881 \\
wa-amazon & temporal & Centroid (embed.) & commodity & 71.7 & 0.32 & 29.6 & 73.3 & 74,881 \\
wa-amazon & temporal & DistilBERT (fine-tuned) & commodity & 84.9 & 0.26 & 31.0 & 84.8 & 74,881 \\
\bottomrule
\end{tabular}
\caption{Complete results grid, part 7 of 8: \textsc{wa-amazon} under \textsc{temporal}. Columns as in Table~\ref{tab:full}.}
\label{tab:full-wa-amazon-temporal}
\end{table*}

\begin{table*}[t]
\centering\footnotesize
\begin{tabular}{lllrrrrrr}
\toprule
Corpus & Protocol & Model & Level & Acc & $\pm$95\% & Macro-$F_1$ & Spend-wtd & $n_\text{test}$ \\
\midrule
wa-amazon & supplier & Majority & segment & 9.9 & 0.13 & 0.4 & 15.3 & 208,845 \\
wa-amazon & supplier & Naive Bayes & segment & 80.3 & 0.17 & 43.6 & 77.8 & 208,845 \\
wa-amazon & supplier & Centroid & segment & 67.5 & 0.20 & 35.2 & 65.2 & 208,845 \\
wa-amazon & supplier & 1-NN & segment & 79.6 & 0.17 & 49.6 & 78.6 & 208,845 \\
wa-amazon & supplier & 1-NN (embed.) & segment & 81.4 & 0.17 & 51.9 & 80.2 & 208,845 \\
wa-amazon & supplier & Centroid (embed.) & segment & 60.2 & 0.21 & 30.8 & 59.4 & 208,845 \\
wa-amazon & supplier & Majority & family & 8.6 & 0.12 & 0.1 & 14.4 & 200,366 \\
wa-amazon & supplier & Naive Bayes & family & 75.5 & 0.19 & 24.3 & 73.1 & 200,366 \\
wa-amazon & supplier & Centroid & family & 60.4 & 0.21 & 24.0 & 57.3 & 200,366 \\
wa-amazon & supplier & 1-NN & family & 75.9 & 0.19 & 33.7 & 76.0 & 200,366 \\
wa-amazon & supplier & 1-NN (embed.) & family & 78.0 & 0.18 & 36.7 & 78.1 & 200,366 \\
wa-amazon & supplier & Centroid (embed.) & family & 56.0 & 0.22 & 22.8 & 53.7 & 200,366 \\
wa-amazon & supplier & Majority & class & 3.0 & 0.08 & 0.0 & 2.1 & 178,930 \\
wa-amazon & supplier & Naive Bayes & class & 75.0 & 0.20 & 20.1 & 72.9 & 178,930 \\
wa-amazon & supplier & Centroid & class & 62.7 & 0.22 & 22.1 & 61.6 & 178,930 \\
wa-amazon & supplier & 1-NN & class & 77.0 & 0.19 & 31.2 & 76.7 & 178,930 \\
wa-amazon & supplier & 1-NN (embed.) & class & 79.2 & 0.19 & 33.6 & 78.3 & 178,930 \\
wa-amazon & supplier & Centroid (embed.) & class & 62.7 & 0.22 & 23.6 & 61.5 & 178,930 \\
wa-amazon & supplier & Majority & commodity & 2.5 & 0.10 & 0.0 & 1.3 & 89,257 \\
wa-amazon & supplier & Naive Bayes & commodity & 77.8 & 0.27 & 20.9 & 77.0 & 89,257 \\
wa-amazon & supplier & Centroid & commodity & 73.7 & 0.29 & 28.9 & 73.0 & 89,257 \\
wa-amazon & supplier & 1-NN & commodity & 83.5 & 0.24 & 38.9 & 83.2 & 89,257 \\
wa-amazon & supplier & 1-NN (embed.) & commodity & 85.0 & 0.23 & 42.8 & 85.6 & 89,257 \\
wa-amazon & supplier & Centroid (embed.) & commodity & 74.5 & 0.29 & 34.4 & 73.0 & 89,257 \\
wa-amazon & supplier & DistilBERT (fine-tuned) & commodity & 85.8 & 0.23 & 29.8 & 85.3 & 89,257 \\
\bottomrule
\end{tabular}
\caption{Complete results grid, part 8 of 8: \textsc{wa-amazon} under \textsc{supplier}. Columns as in Table~\ref{tab:full}.}
\label{tab:full-wa-amazon-supplier}
\end{table*}

\end{document}